\documentclass{aa}  

\usepackage{graphicx}
\usepackage{txfonts}
\usepackage{url}
\usepackage{hyperref}
\hypersetup{
    colorlinks=true,
    linkcolor=blue,
    citecolor=blue,
    urlcolor=blue
    }
\usepackage{pdflscape}
\usepackage{adjustbox}
\usepackage{CJKutf8}
\usepackage{placeins}

\newcommand{\teff}{T_\mathrm{eff}}
\newcommand{\logg}{\log{g}}
\newcommand{\loggf}{\log{gf}}
\newcommand{\vmic}{V_\mathrm{mic}}
\newcommand{\vmac}{V_\mathrm{mac}}

\newcommand{\vsini}{v\sin{i}}
\newcommand{\lv}{\lambda_\mathrm{v}}
\newcommand{\dc}{d_\mathrm{c}}
\newcommand{\pysme}{\texttt{PySME}\xspace}
\newcommand{\sme}{\texttt{SMElib}\xspace}

\begin{document}

  \title{\pysme v1.0: improved modelling of stellar spectra \\for survey-scale applications}

   \author{Mingjie Jian \begin{CJK*}{UTF8}{gbsn}(简明杰)\end{CJK*}\inst{1, 2}
          \and
          Nikolai Piskunov\inst{3}
          \and 
          Jeff Valenti\inst{4}
          \and
          Ella Xi Wang\inst{1}
          \and
          Brian Thorsbro\inst{5,6}
          \and 
          Henrik J\"onsson\inst{7}
          \and 
          Ansgar Wehrhahn\inst{3}
          }

    \institute{Department of Astronomy, Stockholm University, AlbaNova University Center, Roslagstullsbacken 21, 114 21 Stockholm, Sweden\\
              \email{jian-mingjie@outlook.com}
         \and
             Institute of Astronomy, University of Cambridge, Madingley Road, Cambridge CB3 0HA, UK
         \and
             Department of Physics and Astronomy, Uppsala University, Box 516, SE-751 20 Uppsala, Sweden
         \and
            Space Telescope Science Institute, 3700 San Martin Drive, Baltimore, MD 21218, USA
         \and
            Universit\'e C\^ote d’Azur, Observatoire de la C\^ote d’Azur, CNRS, Laboratoire Lagrange, 06000 Nice, France
         \and
            Division of Astrophysics, Department of Physics, Lund University, Box 118, SE-22100 Lund, Sweden
         \and
            Materials Science and Applied Mathematics, Malm\"o University, SE-205 06 Malm\"o, Sweden
             }
    \titlerunning{\pysme v1.0}
    \authorrunning{M. Jian et al.}
   \date{Received March 23, 2026; accepted May 04, 2026}

 
  \abstract
   {
    Stellar abundance analysis relies on flexible, high-performance spectral synthesis.
    To meet these needs, we present \pysme\ v1.0, an updated Python implementation of Spectroscopy Made Easy (SME) designed for precise and survey-scale modelling of stellar spectra.
    A central challenge in SME based synthesis is the efficient treatment of very large line lists, including both the preselection of negligible lines and the subsequent formal synthesis.
    \pysme\ v1.0 introduces a revised line-selection framework based on opacity ratio and line depth, together with dynamic line list construction and control of the effective wavelength span over which each line contributes to the synthetic spectrum.
    These workflows support parallel preprocessing of weak-line selection and reduce the line list passed to the synthesis core, thereby improving scalability while preserving synthetic accuracy.
    \pysme\ v1.0 also incorporates an updated equation-of-state treatment that improves the modelling of hydrogen lines, particularly Balmer features, while maintaining close agreement with previous SME results for metal lines.
    The Python interface has further been extended to support parameter-dependent derived quantities updated during optimisation, and \pysme\ provides non-local thermodynamic equilibrium (NLTE) departure-coefficient grids for 17 elements.
    Together, these developments establish \pysme\ v1.0 as a robust and efficient framework for high-precision stellar abundance analyses in large spectroscopic surveys.

   }


   \maketitle
%

\section{Introduction}
\label{sec:intro}

Analysing stellar spectra is a cornerstone of modern astronomy, driving research in fields such as exoplanets, stellar evolution, and particularly galactic archaeology. 
Galactic archaeology seeks to reconstruct the formation and evolution of our Galaxy by extracting dynamical and chemical information from its constituent stars, much of which is encoded in their spectra (see e.g., \citealt{Tolstoy2011}). 
From these spectra, absorption line shifts yield line-of-sight stellar velocities, while line strengths diagnose physical conditions in the photosphere, such as temperature, surface gravity, and elemental abundances.
However, abundances themselves cannot be directly read off from the spectra.
Instead, they are usually inferred by comparing observed spectra with synthetic models generated under specific assumptions. 
This process involves iteratively adjusting stellar parameters until the synthetic spectrum matches the observed one within acceptable uncertainties, or (alternatively) deriving abundances from measured equivalent widths by matching them to model predictions (see \citealt{Jofre2019} and references therein).
Consequently, accurate and efficient modelling of radiative transfer in stellar atmospheres is essential for reliable spectroscopic analysis.
This need is particularly acute in the era of large Galactic archaeology surveys, including LAMOST \citep{Cui2012, Deng2012}, APOGEE \citep{Majewski2017}, Gaia-ESO \citep{Gilmore2012, Randich2013}, GALAH \citep{DeSilva2015, Buder2021}, WEAVE \citep{Dalton2012, Jin2024}, and 4MOST \citep{deJong2019}, which provide or will soon provide spectra for large stellar samples across the Milky Way. 

A number of radiative transfer codes have been developed for stellar spectral synthesis, including SYNTHE \citep{Kurucz1993}, SME \citep{Valenti1996}, SYNSPEC \citep{Hubeny2011}, MOOG \citep{Sneden2012}, Turbospectrum \citep{Plez2012, Gerber2023}, and Korg \citep{Wheeler2023}.
These tools take a stellar atmospheric structure and a line list as inputs, and predict the emergent spectrum by solving the radiative-transfer equation under local thermodynamic equilibrium (LTE) or non-LTE (NLTE) assumptions. 
In practice, the first step is to use the equation of state (including temperature-dependent partition functions) to determine chemical equilibrium, ionisation equilibrium, and excitation populations.
Next, one evaluates the line opacity (accounting for intrinsic line broadening, e.g., Doppler, collisional, and microturbulence), the continuous opacity, and the source function as a function of wavelength and depth in the atmosphere.
Finally, radiative transfer along a sight line yields an emergent intensity spectrum that is convolved with broadening kernels that represent macroturbulence, rotation, and the instrumental line spread function.

Spectroscopy Made Easy, or SME \citep{Valenti1996}, is a widely used framework for forward modelling and fitting high-resolution spectra, designed to provide an easy-to-use interface with built-in support for line data from The Vienna Atomic Line Database (VALD3; \citealt{Piskunov1995, Ryabchikova2015}).
In its original release, SME used a compiled library (\texttt{SMElib}) to compute ionisation balance, continuous opacity, and line opacity with classical pressure broadening.
The library solved radiative transfer for multiple sight lines, assuming local thermodynamic equilibrium (LTE). 
An interface written in the Interactive Data Language (IDL) interpolated model atmospheres from an ATLAS9 grid \citep{Kurucz1993b} in effective temperature and gravity, integrated specific intensities over the stellar disk to obtain a flux spectrum, and optionally fitted an observed spectrum.

Subsequent development described in \citet{Piskunov2017} modernised both the physical modelling and numerical implementation. 
A new equation-of-state library was introduced to solve chemical and ionisation equilibrium self-consistently across a wide range of temperatures and pressures. 
The new, efficient radiative-transfer solver based on a B\'ezier spline approximation to the source function can handle both plane-parallel and spherical model geometries.
Additional model atmosphere grids including MARCS \citep{Gustafsson2008} were included in the SME distribution. 
NLTE synthesis is enabled through precomputed departure coefficients for many species.

\pysme \citep{Wehrhahn2023} is an open source Python package for spectral synthesis and fitting that wraps \sme. 
SMElib remains responsible for the performance-critical radiative-transfer and opacity calculations that generate synthetic intensity spectra, while all higher-level functionality is implemented in Python instead of IDL. This dramatically reduces software licensing costs, making survey-scale analyses feasible.
In synthesis mode, the user specifies line data (often directly from VALD), stellar parameters (e.g., $\teff$, $\logg$, abundances), and \pysme produces intensity spectra and usually a disk-integrated flux spectrum. 
\pysme can also determine stellar parameters by fitting an observed spectrum using a least-squares optimiser.
\pysme also manages NLTE departure coefficients by interpolating precomputed grids to the target stellar parameters and passing them to the \sme radiative transfer core.
These design choices preserve the performance and precision of the synthetic spectra from \sme, while improving accessibility, transparency, and extensibility for installation, reproducible workflows, and future development.
For basic usage examples, see the documentation\footnote{\url{https://pysme-astro.readthedocs.io/en/latest/}} and \citet{Wehrhahn2023}.

With the increasing volume of spectroscopic data from current and upcoming large-scale surveys such as LAMOST, 4MOST, and WEAVE, improving performance and scalability of spectroscopic analysis is essential. 
This issue, along with the dominant computational bottleneck of \pysme, is discussed in Section~\ref{sec:computation-cost} and motivates the development of updated versions of \sme and \pysme (Section~\ref{sec:new_sme}). 
Building on \pysme\ v1.0, we introduce a line-filtering algorithm in Section~\ref{sec:dynamic_linelist} and benchmark its performance in Section~\ref{sec:result}.

\section{Opacity for all the lines: the major computation cost}
\label{sec:computation-cost}

Stellar spectroscopic data are growing rapidly in resolution, wavelength coverage, and sample size, thanks to improved telescope capabilities and the rise of large-scale spectroscopic surveys. 
This growth places increasing demands on the accuracy and efficiency of spectral analysis, especially in high-resolution spectral synthesis.
A major computational challenge in this context is the vast number of spectral lines to be included, which makes the calculation of line opacities and emergent spectra both time-consuming and resource-intensive.

\sme addresses this demand by combining a weak-line filtering algorithm and an adaptive wavelength grid during synthesis \citep[see][Section 2.3]{Valenti1996}. 
It first computes, for each line, the maximum line-to-continuum opacity ratio at line centre across all atmospheric layers, and discards lines whose maximum ratio falls below a user-defined threshold.
The centres of the retained lines then provide an initial set of wavelength grid nodes.
Starting from this grid, \sme iteratively refines the wavelength sampling by inserting midpoints between neighbouring nodes until the interpolation error in the specific intensity drops below a prescribed tolerance.
As a result, the wavelength grid becomes dense where the intensity varies rapidly and remains sparse where it changes slowly.
Finally, the emergent flux is obtained by disk-integrating the computed intensities and is then linearly interpolated to the user-defined wavelength grid.
By explicitly controlling the interpolation error while preventing negligible lines from driving unnecessary grid refinement, this approach maintains accuracy while reducing computational cost.

The current implementation has two main limitations.
A simple cut based on a fixed threshold may inadvertently remove groups of individually weak lines that collectively contribute to observable spectral features (e.g. molecular bands), while a criterion formulated in terms of an opacity-ratio threshold does not make it easy to predict the impact of a chosen threshold on the synthetic spectrum.
For very large line lists, the weak line screening can itself become expensive when performed serially, and the efficiency of the subsequent synthesis also depends on how the retained lines are traversed and incorporated into the opacity calculation.
\pysme therefore adopts a depth-like proxy (line depth) as an alternative line-selection metric, which connects the threshold more directly to the output spectrum.
To improve scalability, \pysme moves line preselection to a parallel preprocessing stage, supports reuse of precomputed line information, and reduces the effective line list passed to the synthesis core.
This design enables both faster repeated optimisation runs and more interpretable control of line-selection thresholds.

\section{\pysme v1.0}
\label{sec:new_sme}

To address the computational bottlenecks associated with large line lists, \pysme\ v1.0\footnote{\url{https://github.com/SpectroscopyMadeEasy/PySME}} introduces a revised line-selection framework for spectral synthesis and fitting.
The key developments are a unified interface for line preselection, support for parallel preprocessing of line information, and improved integration with \sme\ so that the synthesis core can make more efficient use of the selected lines.
In addition, \pysme\ v1.0 introduces a contribution-function utility for analysing line-formation regions and incorporates an updated equation-of-state treatment in \sme.
These updates are described in the following subsections.

\subsection{Line diagnostics from \sme}
\label{sec:line_cdvr}

PySME uses several per-line diagnostics from \sme\ to support line selection, including the central depth ($d_\mathrm{c}$), the maximum line-centre opacity ratio (ALMAX), and the validity range ($\lambda_\mathrm{v}$).
These quantities are computed internally by \sme\ and have been exposed to the Python layer since \pysme\ v0.6.

Central depth is the depth at the centre wavelength ($\lambda_0$) of a line before applying macroturbulent ($\vmac$), rotational ($\vsini$), or instrumental broadening, i.e.
\begin{equation}
    d_\mathrm{c} = 1 - \frac{F^{\mathrm{line+cont}}_{\lambda_0}}{F^{\mathrm{cont}}_{\lambda_0}},
\end{equation}
where \(F^{\mathrm{line+cont}}_{\lambda_0}\) is the disk-integrated emergent flux computed at \(\lambda_0\) with both line and continuum opacity, and \(F^{\mathrm{cont}}_{\lambda_0}\) is the corresponding continuum-only flux.
Here, intrinsic line broadening refers to the combined effects of thermal Doppler broadening, natural broadening, and pressure broadening (including Stark and van der Waals effects).
Because observed spectra additionally include broadening from $\vmac$, $\vsini$, and the instrumental profile, the observed line-centre depth is generally shallower than $d_\mathrm{c}$.
These additional broadening mechanisms are not part of the definition of $d_\mathrm{c}$; their implementation in \sme\ and \pysme, including the rotational, macroturbulent, and instrumental kernels, is described by \citet{Valenti1996} and \citet{Wehrhahn2023}.
This makes $d_\mathrm{c}$ a natural and conservative quantity for threshold-based line selection.

As an alternative diagnostic, \sme\ also provides the maximum line-to-continuum opacity ratio at line centre across atmospheric layers (hereafter ALMAX), which serves as the default line-selection algorithm.
Compared with the central depth, ALMAX is computationally lighter to evaluate, whereas the central depth is more directly interpretable in terms of the impact on the output spectrum. 
In this work, we therefore use the central depth as the main quantity for describing and controlling line-selection behaviour, while retaining ALMAX as a faster alternative.

In addition to line strength, it is also necessary to determine the wavelength interval over which a line contributes significantly to the opacity.
For example, synthesising a spectrum using only lines whose centres fall between 3969 and 3970\,\AA{} would neglect the extended wing of the strong \ion{Ca}{ii} H line centred outside this interval.
In \sme, each line is therefore associated with a validity range, defined as the smallest symmetric interval around $\lambda_0$ within which the line-to-continuum opacity ratio remains above a user-set threshold, excluding the effects of $\vmac$, $\vsini$, and instrumental broadening.
To determine this range, \sme\ evaluates the opacity ratio at $\lambda_0+\Delta\lambda$ for increasing values of $\Delta\lambda$, starting from 0.3\,\AA{}, until the ratio falls below the threshold set by \texttt{accrt} (default 0.0001).
The resulting interval is adopted as the validity range ($\lambda_\mathrm{v}$) for that line.

The diagnostics $d_\mathrm{c}$, ALMAX, and $\lambda_\mathrm{v}$ provide complementary information about line strength and wavelength extent.
PySME uses them in different ways depending on the selected line-selection method; the corresponding workflows are described in the following subsection.

\subsection{Unified line-selection workflows in \pysme}
\label{sec:line_select_workflow}

\pysme\ v1.0 provides a unified interface for line selection, allowing the user to choose among three workflows: \texttt{internal}, \texttt{almax}, and \texttt{cdr}.
The \texttt{internal} workflow retains the legacy behaviour of \sme, in which weak-line screening and validity-range determination are performed inside the synthesis core during spectrum calculation.
The \texttt{almax} and \texttt{cdr} workflows instead precompute per-line selection information in Python and pass it to \sme\ before synthesis.

The \texttt{almax} workflow performs line selection using ALMAX together with the validity range.
Because ALMAX is relatively inexpensive to evaluate, this workflow provides an efficient preselection strategy and is well suited to large line lists.
The \texttt{cdr} workflow performs line selection using the central depth together with the validity range.
Although the central depth is more expensive to compute, it is more directly interpretable in terms of the impact on the output spectrum, and is therefore used in this work as the main quantity for describing and controlling line-selection behaviour.

These workflows are controlled through a common configuration interface.
The parameter \texttt{line\_select\_method} chooses the backend (\texttt{internal}, \texttt{almax}, or \texttt{cdr}), while additional options control the associated workflow, including whether preselection is parallelised, how many worker processes are used, and whether previously computed line information should be recomputed or reused.
This design makes it possible to switch between line-selection strategies without changing the downstream synthesis workflow.

When precomputed line information is available, \pysme\ passes the corresponding line masks and validity ranges to \sme\ through its line-information interface before synthesis.
This allows the synthesis core to bypass redundant line-selection work and to make more efficient use of the selected lines during opacity calculation.
Further details on reuse, caching, and dynamic line list construction are given in Section~\ref{sec:dynamic_linelist}.

\subsection{Contribution function}
\label{sec:contrubition_func}

Spectral lines are formed in specific regions of stellar atmospheres, rather than throughout the entire atmosphere. 
Their strength and profile are closely linked to the local physical conditions within these formation regions. 
Typically, the line core originates from higher layers in the photosphere, while the wings are shaped by deeper layers. 
It is therefore useful to characterise the effective line formation region to understand which atmospheric layers primarily shape the line core and wings, and hence which physical conditions the line is sensitive to.

The contribution function describes how much light each atmospheric layer contributes to the emergent radiation at a given wavelength \citep[see][Chapter 7]{Gray2008}. 
In other words, it identifies the depth regions that dominate the formation of intensity or flux at each wavelength point.
Starting with version~v1.0, \pysme provides the contribution function of the synthetic spectrum as an optional output. 
It is returned as an $N\times M$ array, where 
$N$ is the number of wavelength points and 
$M$ is the number of atmospheric layers, allowing the function to be evaluated across the entire spectrum.

Other forms of contribution functions also exist, such as those that quantify the total absorption produced in each layer \citep[e.g.][]{Magain1986, Achmad1992, Gurtovenko1991} for the 1D atmosphere case. 
Support for the calculation of these absorption-type contribution functions as optional output is planned for future releases of \pysme.

\subsection{Modifications in \sme}

In \texttt{PySME} v1.0, we update the bundled \texttt{SMElib} to v6.13 and apply a small set of targeted changes relative to earlier releases. 
These updates include both physics-motivated improvements to the underlying calculations and implementation-level changes that make the synthesis workflow more efficient when precomputed line information is available.
We revised the data stored in the EOS solver to resolve the problem of consistency of thermodynamic quantities (in particular, the EOS-computed electron density) with the input model atmosphere. The problem was traced to the new set of partition functions for tri-atomic molecules that was misinterpreted by the solver. 
Although this problem had negligible impact on most metal lines at temperatures above 5000~K, it could lead to noticeable differences in line strength and the wings of Balmer lines for cooler temperatures.  
This problem is now fixed.

In addition, the EOS chemical network was extended to new molecular species/ions (e.g., NH$_2$) and negative atomic ions to ensure a more complete equilibrium calculation. 
We also updated the \ion{Fe}{i} photoionisation cross-section table used for continuous-opacity calculations from \citet{Bautista2017} to \citet{Zatsarinny2019}. 
We therefore recommend using the updated \texttt{SMElib} (and thus \texttt{PySME} v1.0) rather than earlier releases.
A dedicated comparison of the old and new \texttt{SMElib} is provided in Section~\ref{sec:SMElib_comp}.

\subsection{Miscellaneous}
\label{sec:miscellaneous}

In addition to the core updates described above, \pysme v1.0 also introduces several auxiliary improvements that enhance its flexibility and usability in practical applications. 
These include support for derived parameters during fitting and expanded NLTE grid coverage, as detailed below.

\subsubsection{Derived parameters}

\pysme now allows some parameter (named as ``derived parameters'' and \texttt{derived\_param} in the package) changes following the variation of fitting parameter during fitting. 
Some of the stellar parameters are physically correlated due to stellar structure and/or evolution.
For example, $\teff$ and $\logg$ are positively correlated along the red-giant branch; surface C, N, and O abundances evolve as stars ascend from dwarfs to giants; and $[\alpha/\mathrm{Fe}]$ abundance ratios tend to be higher at low metallicity.
While these correlations reflect underlying physical processes, they do not necessarily imply degeneracies in spectral fitting.
In such cases, certain quantities may be better treated as derived parameters mainly to accelerate the fitting process by constraining correlated quantities through physically motivated relationships.

During the $\chi^2$ minimisation, the derived parameters are recomputed after each iteration based on a user-defined function from the current fitted parameters, and the updated values are used in the next fitting step.
This approach has been shown to reduce the number of iterations in large surveys such as the GALAH survey and to improve results in star-based fitting \citep{Bijavara2025}.

It is important to note that treating quantities as derived can affect uncertainty estimates from $\chi^2$ minimisation. 
If a derived parameter strongly influences the spectrum, the standard error bars derived in this case may be biased. 
We therefore recommend using this option with care and verifying uncertainties when such effects are significant.

\subsubsection{Updated NLTE grids}

\begin{figure*}[!t]
    \centering
    \includegraphics[width=1\linewidth]{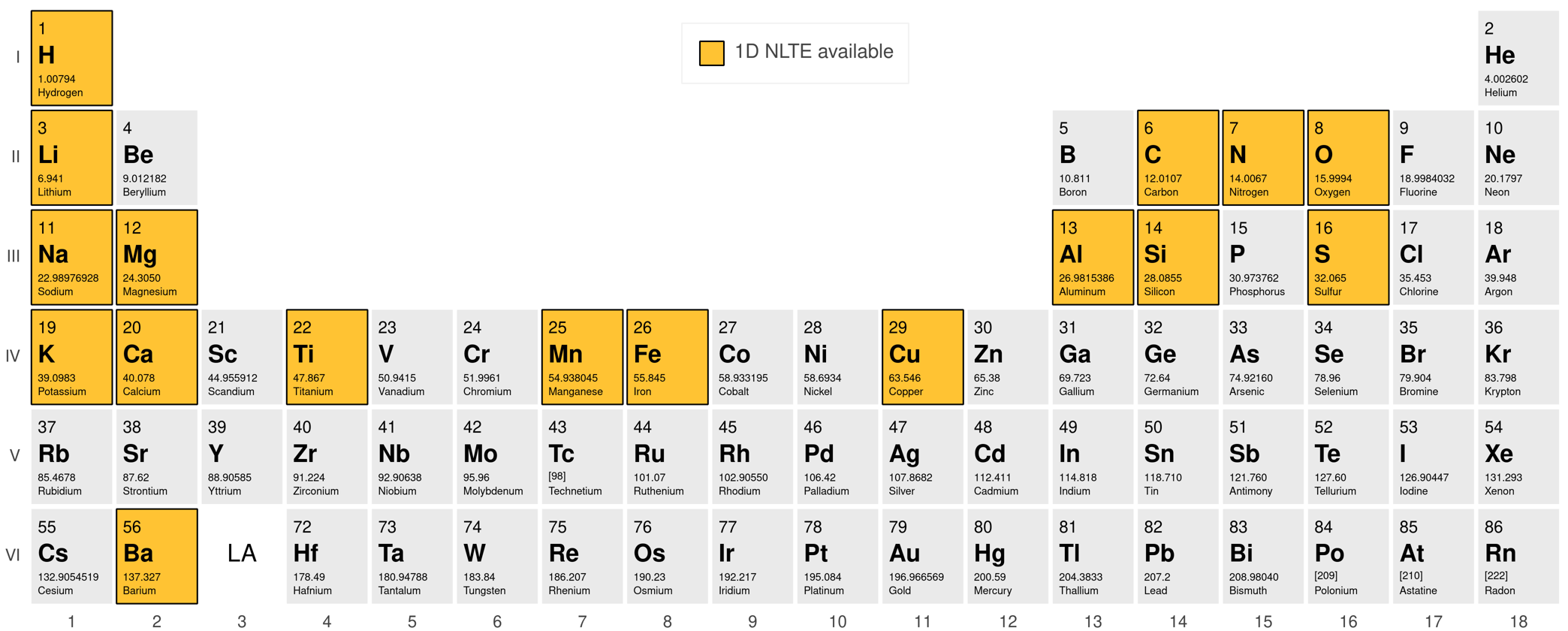}
    \caption{Periodic table highlighting the chemical elements for which 1D NLTE departure coefficients are available in \pysme.}
    \label{fig:nlte_table}
\end{figure*}

Using departure coefficients to approximate NLTE effects is supported in \sme \citep{Piskunov2017}.
Through \pysme, precomputed grids of departure coefficients for the MARCS grid can be loaded and interpolated to provide the inputs required by \sme. 
In total, NLTE grids are available for 17 elements\footnote{\url{https://zenodo.org/records/17064337}} (Figure~\ref{fig:nlte_table}, \citealt{Amarsi2020}); three elements, S, Ti, and Cu have been added since \citet{Wehrhahn2023}. 
References for the model atoms used for each element are listed in Table~\ref{tab:NLTE-grids}.
The NLTE grids follow the MARCS model-atmosphere grid points.
The available abundance ranges are as follows: for the grids described by \citet{Amarsi2020}, a generic element A is generally tabulated over $-2 \le [\mathrm{A/Fe}] \le +2$, with H tied to the model-atmosphere composition and Li covering $-4 \le [\mathrm{Li/Fe}] \le +7$.
The S, Ti, and Cu grids cover $-1.5 \le [\mathrm{S/Fe}] \le +1.5$, $-2 \le [\mathrm{Ti/Fe}] \le +2$, and $-1.2 \le [\mathrm{Cu/Fe}] \le +1.2$, respectively.
The grids should therefore only be used within their available parameter and abundance coverage.

\section{Dynamic line list in \pysme}
\label{sec:dynamic_linelist}

For the Python side line-selection workflows in \pysme, the line list passed to \sme\ is constructed dynamically from precomputed line information rather than from the full input list.
After computing the relevant line diagnostics and validity ranges for all lines, \pysme\ filters out negligible lines according to the algorithm described in Section~\ref{sec:negligible-line-filtering}; these diagnostics can be obtained either serially or in parallel depending on the selected workflow.
It then further excludes any line whose validity range does not overlap the coverage of at least one synthesis segment.
As a result, the synthesis step avoids processing both negligible lines and lines that lie entirely outside the relevant wavelength segments, thereby reducing the computational cost.

Unlike previous versions of \pysme, where all input lines were passed to \sme\ regardless of their significance, the dynamic line list contains only the subset of lines deemed relevant for the current synthesis setup.
The line list is therefore “dynamic” in the sense that it depends on the selected workflow, the current line-selection criteria, and the wavelength segments being synthesised.

\subsection{Accumulated negligible-line filtering}
\label{sec:negligible-line-filtering}

For the Python-side line-selection workflows, \pysme\ supports an accumulated negligible-line filtering strategy based on a per-line diagnostic $q$, which may be either the central depth ($\dc$) or ALMAX.
A negligible line is then defined relative to a user-defined threshold $q_\mathrm{w}$.

Naively discarding all lines with $q \le q_\mathrm{w}$ does not guarantee that the final synthetic spectrum remains unaffected at the same level.
A cluster of individually weak lines can still produce a combined absorption feature if they are crowded in wavelength (e.g. TiO lines in the optical).
The same consideration applies to hyperfine-structure (HFS) components when they are provided as separate nearby transitions in the input line list: the accumulated filtering treats them like any other crowded line groups and retains them when their combined local contribution exceeds the threshold.
To prevent this, \pysme\ groups spectral lines into short wavelength bins (0.2\,\AA{} by default), sorts the lines in each bin by the chosen diagnostic $q$, and discards the weakest lines until the accumulated value reaches $q_\mathrm{w}$.
This procedure ensures that both isolated strong lines and clusters of weak lines are retained whenever their combined contribution may become significant within a local wavelength interval.

When $q=d_\mathrm{c}$, the threshold is more directly interpretable in terms of the impact on the continuum-normalised spectrum, and $q_\mathrm{w}$ corresponds to the depth threshold $d_\mathrm{w}$ used throughout this work.
When $q=\mathrm{ALMAX}$, the same accumulated filtering strategy can be applied, but the threshold is no longer directly interpretable as a flux-depth limit in the output spectrum.
The choice of a 0.2\,\AA{} bin width matches the typical width of a weak line; users may adjust this value if a different resolving power or line-width regime is relevant.

In summary, the accumulated filtering strategy improves computational efficiency by removing locally negligible lines while retaining features whose combined contribution may still be important for the synthetic spectrum.
In the \texttt{internal} workflow, line selection follows the original threshold-based filtering implemented in \sme.
For the Python-side \texttt{almax} and \texttt{cdr} workflows, \pysme\ instead applies the accumulated filtering strategy described in Section~\ref{sec:negligible-line-filtering}.

\subsection{Line-information database}

When performing spectral synthesis for large samples or repeated optimisation runs, the input line list is usually fixed.
For the Python-side line-selection workflows in \pysme, the per-line diagnostics and validity ranges associated with a given set of stellar parameters can therefore be reused once they have been computed.
To facilitate this, \pysme\ provides a database-backed mechanism for storing and reusing line information.

Depending on the selected workflow, the stored diagnostic is either the central depth (\(d_\mathrm{c}\)) or ALMAX, together with the corresponding validity range (\(\lambda_\mathrm{v}\)) and the stellar parameters (\(\teff\), \(\logg\), and metallicity).
When a target directory is specified, \pysme\ stores this information in \texttt{.npz} files and appends new grid points on demand.
The database therefore grows lazily during use and does not require a fully precomputed stellar-parameter grid in advance.

On subsequent calls, \pysme\ first checks whether the target parameters coincide with, or lie sufficiently close to, a previously stored grid point.
The corresponding line information is then reused directly, without recomputing the diagnostics from scratch.
If no such nearby grid point is available, \pysme attempts the following.
When the target parameters fall within a simplex of the precomputed grid and the surrounding grid points satisfy user-defined tolerances, \pysme derives the negligible-line mask and validity range from the neighbouring grid points and uses them to construct the dynamic line list.
Otherwise, the required diagnostic and validity range are computed for the target parameters and added to the database as a new grid point.

Direct interpolation of the raw line diagnostic can degrade precision, especially when the stellar-parameter grid is sparse.
For this reason, the database workflow is designed primarily for conservative reuse of line-selection information rather than for high-accuracy interpolation of the diagnostic itself.
In practice, a line is retained at the target parameters if it is classified as non-negligible at any of the neighbouring grid points contributing to the local simplex.
The corresponding validity range, however, is still obtained by linear interpolation from the same neighbouring grid points.
This preserves the conservative nature of the line-selection step while avoiding unnecessary recomputation.

Taken together, the accumulated filtering strategy and database-backed reuse provide the main Python-side mechanisms for constructing the dynamic line list in \pysme.

\section{Result and validation}
\label{sec:result}

We now turn to the validation of the methods introduced in Sections~\ref{sec:new_sme} and \ref{sec:dynamic_linelist}. 
This section presents results that illustrate the behaviour of the central depth, validity range, and contribution function, as well as the performance of the negligible-line filtering algorithm. 
We begin by examining individual spectral lines to show how these diagnostics behave under different stellar conditions.  
The subsequent subsections evaluate the filtering algorithm for Sun and Arcturus and across the Kiel diagram, compare synthetic spectra generated with the old and new \sme, and assess the time performance improvements enabled by the dynamic line list.

The stellar parameters of the two benchmark stars used here, the Sun and Arcturus, are listed in Table~\ref{tab:stellar-paras} \citep{Blanco-Cuaresma2014}. 
The line list, including atomic and molecular lines, is extracted from the VALD database and spans $3700$--$9500\,\text{\AA}$ with a total of $3{,}706{,}996$ transitions.
HFS components are included as separate transitions where available, using the corresponding option in the VALD database.
Species with hyperfine structure data in VALD3 are listed in the VALD3 documentation\footnote{\url{https://www.astro.uu.se/valdwiki/VALD3linelists\#Hyperfine_structure_data}}.

\begin{table*}
    \centering
    \caption{Stellar parameters adopted for the Sun and Arcturus.}
    \begin{tabular}{clrrccc}
    \hline
    Star name & $\teff$ & $\logg$ & [Fe/H] & $\vmic$ & $\vmac$ & $\vsini$\\
    & (K) & & & ($\mathrm{km\,s^{-1}}$) & ($\mathrm{km\,s^{-1}}$) & ($\mathrm{km\,s^{-1}}$) \\
\hline\hline
    Sun & $5771$ & $4.44$ & $0.00$ & $1.0$ & $4.19$ & $1.6$ \\
    Arcturus & $4277$ & $1.58$ & $-0.55$ & $1.43$ & $5.12$ & $1.6$ \\
    \hline
    \end{tabular}
    \label{tab:stellar-paras}
\end{table*}

\subsection{Behaviour of $\dc$, $\lv$, and the contribution function}

Although both ALMAX- and $\dc$-based workflows are available in \pysme, we use $\dc$ here as the main diagnostic for illustrating line-selection behaviour because it is more directly interpretable in the continuum-normalised spectrum.
We begin with two representative \ion{Fe}{i} lines, examining their central depth, validity range, and contribution function to illustrate their behaviour.

\subsubsection{Two example \ion{Fe}{i} lines}

\begin{figure*}[!t]
\resizebox{12cm}{!}{\includegraphics{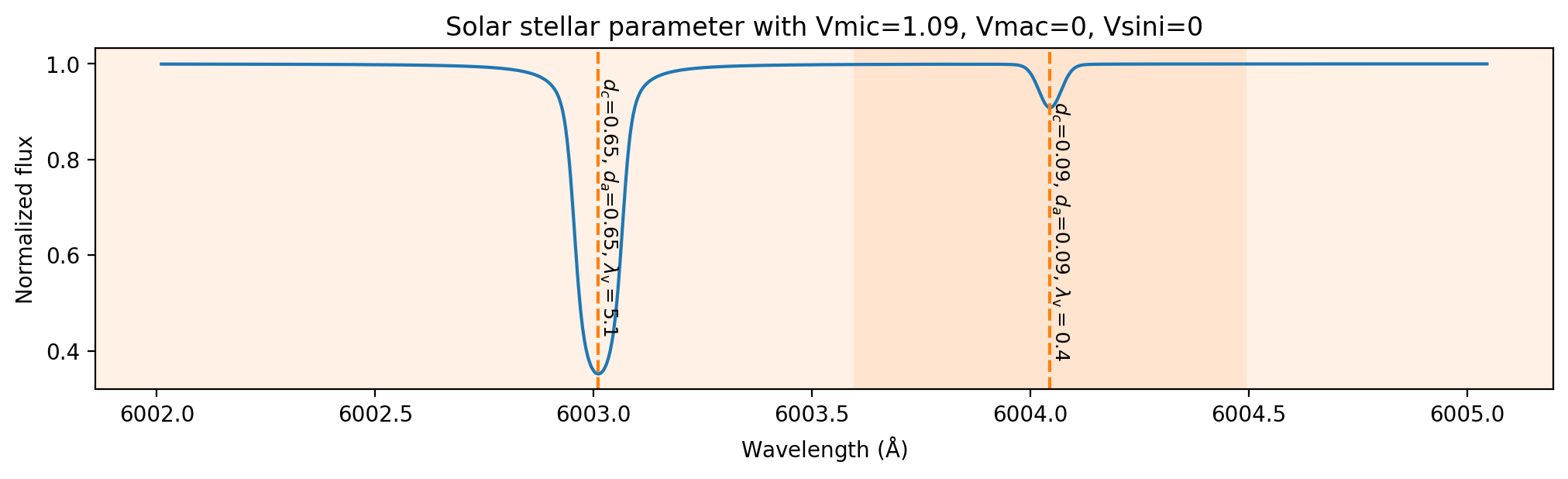}}
\hfill
\parbox[b]{55mm}{
\caption{Synthetic spectrum of the example lines listed in Table~\ref{tab:example_line}, computed using solar stellar parameters. The vertical dashed lines indicate the line centres, with $d_c$, $d_a$, and $\lambda_v$ labelled. The shaded regions show the validity range of each line.}
\label{fig:cdr_example}
}
\end{figure*}

We select one strong and one weak \ion{Fe}{i} line under solar stellar parameters as examples, with their atomic parameters listed in Table~\ref{tab:example_line}.
The synthetic spectrum is computed with no macroturbulence, rotation, or instrumental broadening applied.
Figure~\ref{fig:cdr_example} shows the resulting spectrum, where the two absorption features are clearly visible at 6003.011\,\AA{} and 6004.045\,\AA{}.
In the absence of these broadening mechanisms, the actual depth of the lines, $d_\mathrm{a}$, is equal to $\dc$, confirming that $\dc$ accurately represents the intrinsic line strength.

\begin{table*}
    \centering
    \caption{Example \ion{Fe}{i} lines.}
    \label{tab:example_line}
    \footnotesize
    \begin{tabular}{ccccccccccccc}
    \hline
    species & $\lambda_0$ & $\log gf$ & EP & $j_\mathrm{lo}$ & $E_\mathrm{up}$ & $j_\mathrm{up}$ 
    & lande$_\mathrm{lower}$ & lande$_\mathrm{upper}$ & lande 
    & gamrad & gamqst & gamvw \\
    & (\AA) & & (eV) & & (eV) & 
    & & & & & & \\
    \hline\hline
    \ion{Fe}{i} & $6003.011$ & $-1.120$ & $3.8816$ & $4$ & $5.9464$ & $4$ 
    & $1.25$ & $1.29$ & $1.27$ & $7.75$ & $-5.38$ & $898.241$ \\
    \ion{Fe}{i} & $6004.045$ & $-2.549$ & $4.2562$ & $3$ & $6.3207$ & $4$ 
    & $1.24$ & $1.24$ & $1.23$ & $8.42$ & $-4.89$ & $-7.550$ \\
    \hline
    \end{tabular}
    \tablefoot{
    The table lists the atomic parameters of the example \ion{Fe}{i} lines, including the species and transition wavelength ($\lambda_0$), oscillator strength ($\loggf$), excitation potential (EP), total angular momentum quantum numbers of the lower and upper levels ($j_\mathrm{lo}$ and $j_\mathrm{up}$), Landé g-factors for both levels and their average (lande$_\mathrm{lower}$, lande$_\mathrm{upper}$, and lande), as well as radiative, Stark, and van der Waals damping constants (gamrad, gamqst, gamvw).
    }
\end{table*}

\begin{figure*}[!t]
    \centering
    \includegraphics[width=1\linewidth]{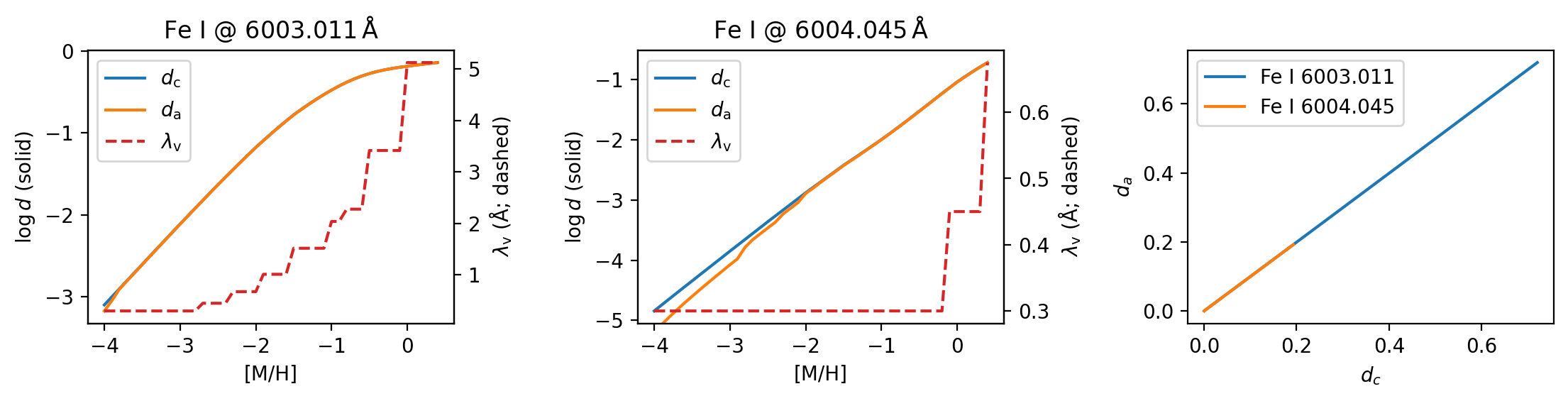}
    \caption{Left and middle panels: growth of $\dc$, $d_\mathrm{a}$ and $\lv$ for two example Fe lines with metallicity in solar parameters. Right panel: $\dc$ vs $d_\mathrm{a}$ for the example Fe lines.}
    \label{fig:dca_plot}
\end{figure*}

Figure~\ref{fig:dca_plot} further illustrates how these parameters vary with metallicity.
Both $\dc$ and $d_\mathrm{a}$ show a ``curve-of-growth'' like behaviour as the metallicity increases from $-4$ to solar. 
For the stronger line, the transition from the linear to the saturated region is clearly visible, whereas the weaker line remains entirely within the linear region across the full metallicity range.
Throughout most of the metallicity range, $\dc$ and $d_\mathrm{a}$ are nearly identical, confirming the reliability of $\dc$ as a proxy for intrinsic line strength.
Minor discrepancies at low metallicity arise from numerical limitations in the synthesis accuracy of \pysme, particularly when line depths are very small.
$\lv$ increases from its initial value of 0.3\,\AA{} for weak lines, and then grows in steps as the line becomes stronger.
At solar metallicity, the validity ranges are 5.1\,\AA{} for the stronger line and 0.4\,\AA{} for the weaker one. 
These ranges extend well beyond the region of significant absorption, highlighting the conservative nature of the threshold-based definition and its sensitivity to line strength.

\begin{figure}[!htbp]
    \centering
    \includegraphics[width=1\linewidth]{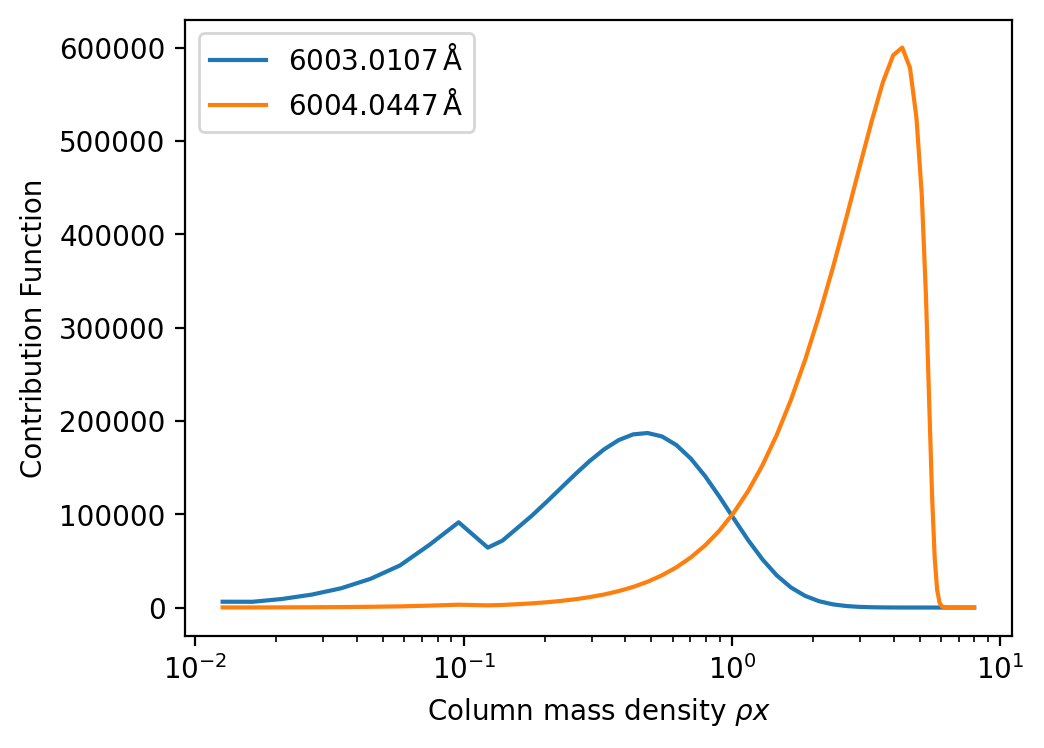}
    \caption{Flux contribution function for the stronger ($6003.011$\,\AA{}) and weaker ($6004.045$\,\AA{}) \ion{Fe}{i} lines, plotted as a function of column mass density $\rho x$. Lower values of $\rho x$ correspond to higher layers in the stellar atmosphere, while higher values represent deeper layers.}
    \label{fig:cf_example}
\end{figure}

Figure~\ref{fig:cf_example} presents the contribution function at the line centres 6003.011\,\AA{} and 6004.045\,\AA{}, computed under solar stellar parameters.
The flux in the stronger line at 6003.011\,\AA{} originates predominantly from higher layers in the stellar atmosphere, while the weaker line at 6004.045\,\AA{} receives its main contribution from deeper regions. 
It reflects the typical formation behaviour of strong versus weak lines: stronger lines tend to form in higher layers, whereas weaker lines probe lower layers.
The apparent decrease of the contribution function at $\rho x \sim 0.1$ for the strong line arises because the atmosphere model samples the depth grid more finely there (smaller $\Delta \rho x$), leading to a smaller $\Delta \tau$ and thus a reduced per-layer integrated contribution.

\subsubsection{Lines in the optical region}

\begin{figure*}[!t]
    \centering
    \includegraphics[width=1\linewidth]{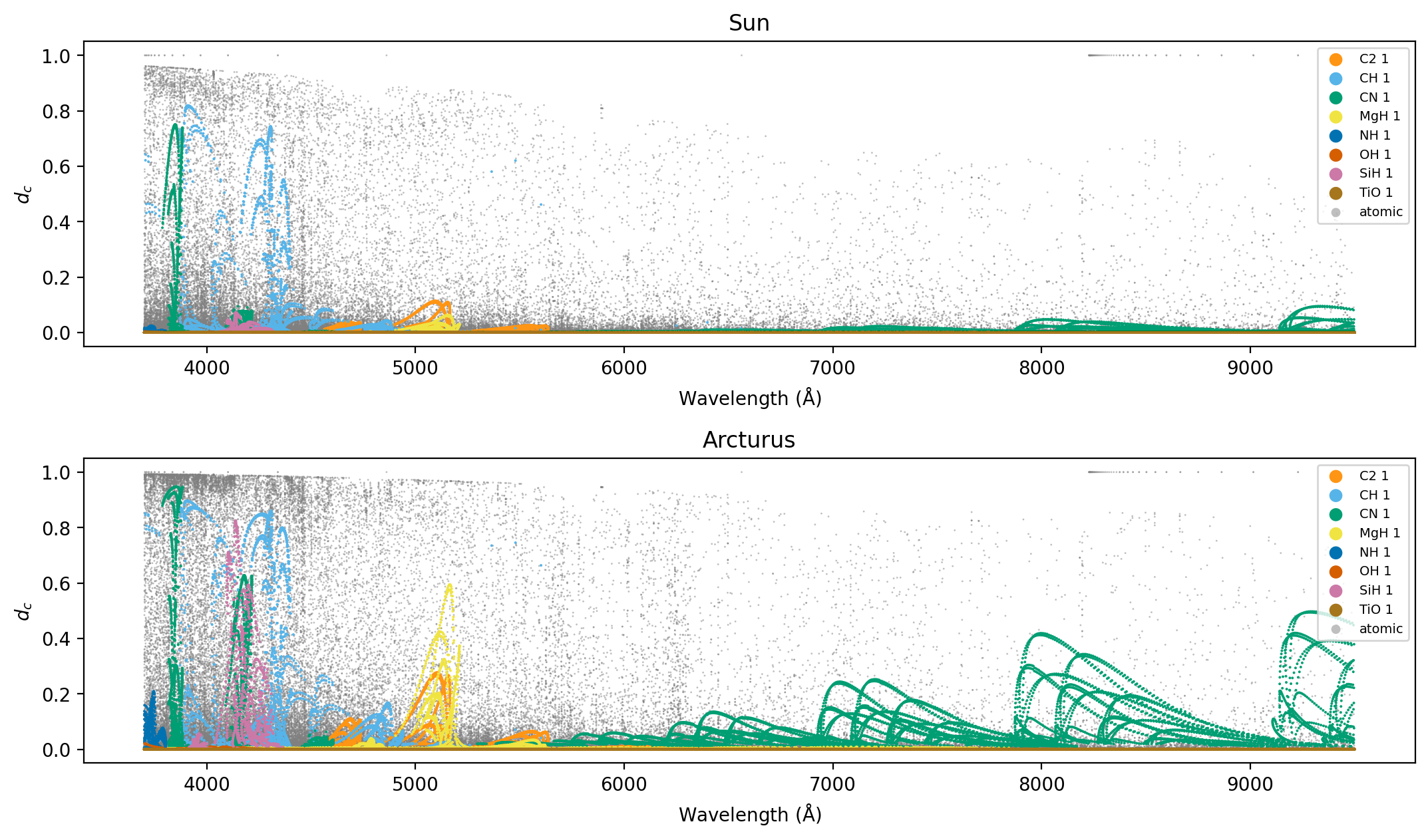}
    \caption{Central depths of optical spectral lines for the Sun (top) and Arcturus (bottom). Coloured points mark molecular features, and grey points show all other atomic lines.}
    \label{fig:sun_arcturus_cd}
\end{figure*}

Figure~\ref{fig:sun_arcturus_cd} shows $\dc$ for all lines in the line list, calculated under solar and Arcturus stellar parameters.
The maximum $\dc$ decreases slightly toward longer wavelengths, likely due to wavelength-dependent continuum opacity (dominated by H$^-$) that lowers the line-to-continuum contrast at longer $\lambda$. 
Lines with $\dc=1$ are exclusively hydrogen lines. 
For molecular features, the central depths are systematically higher in Arcturus than in the Sun. 
Strong CN, MgH, OH, CH, and SiH bands are clearly visible in the Arcturus panel, whereas these molecular features appear significantly weaker in the solar case. 
This contrast arises from the cooler temperature and lower surface gravity of Arcturus, which favour molecule formation and enhance molecular absorption in the optical region. 

\subsection{Negligible-line filtering}

\begin{figure*}[!t]
    \centering
    \includegraphics[width=1\linewidth]{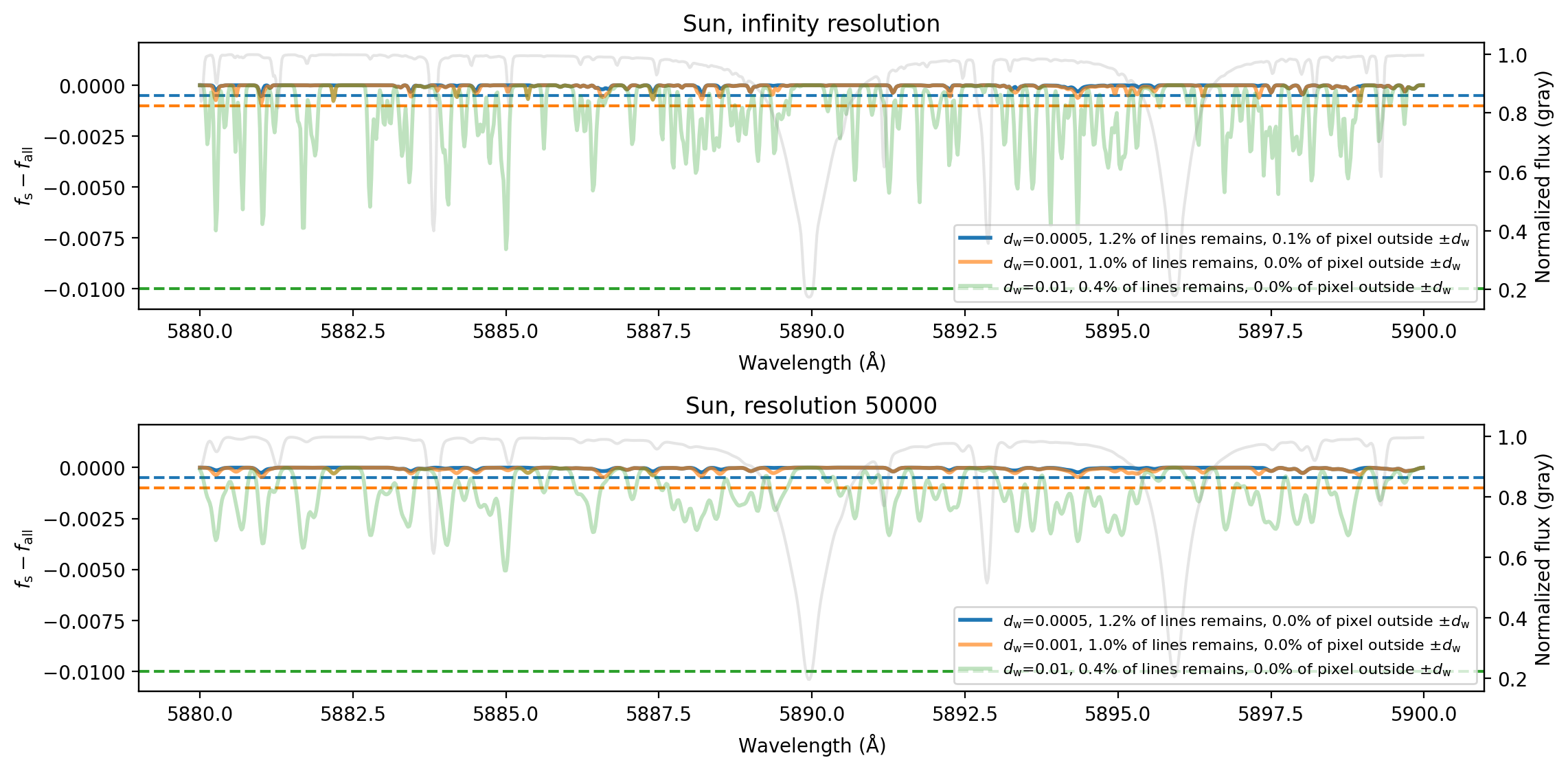}
    \caption{Difference between synthetic spectra computed with negligible lines removed ($f_\mathrm{s}$) and those including all lines ($f_\mathrm{all}$), shown for solar parameters at infinite resolution (top) and resolution 50,000 (bottom). Horizontal dashed lines indicate the $d_\mathrm{w}$ thresholds, and $f_\mathrm{all}$ is plotted in grey as a reference.}
    \label{fig:cd_thres_sun}
\end{figure*}

Figure~\ref{fig:cd_thres_sun} compares synthetic spectra for solar parameters (with $v_\mathrm{mac}=v\sin i=0$, no instrumental broadening, and $\Delta\lambda=0.02$) using the full line list filtered with three different thresholds of $d_\mathrm{w}$, across a wavelength region densely populated by TiO lines.
As expected, spectral differences become more pronounced with increasing $d_\mathrm{w}$, reflecting the progressive exclusion of weak lines.
At infinite resolution and a strict threshold of $d_\mathrm{w} = 0.0005$, fewer than 0.1\% of pixels show deviations exceeding the threshold. 
This confirms that the negligible line filtering algorithm preserves spectral fidelity within the defined tolerance, even under idealised high-resolution conditions.

In practical applications, observed spectra are subject to instrumental broadening, which further decreases the impact of missing weak lines.
The lower panel of Fig.~\ref{fig:cd_thres_sun} shows results at a resolution of $R = 50,000$, where pixel-wise differences are even smaller than in the infinite-resolution case, with no pixel exceeding the $d_\mathrm{w}$ threshold.
This demonstrates that the filtering strategy remains robust under realistic observational conditions, making it well suited for survey-scale synthesis and fitting.
For reference, matching the \(d_\mathrm{w}=5\times10^{-4}\) selection with an ALMAX-based criterion in this solar test requires an empirical threshold of \(\mathrm{ALMAX}\approx1.5\times10^{-3}\), which is substantially larger than the default value of \(10^{-4}\).

\begin{figure*}[!t]
    \centering
    \includegraphics[width=1\linewidth]{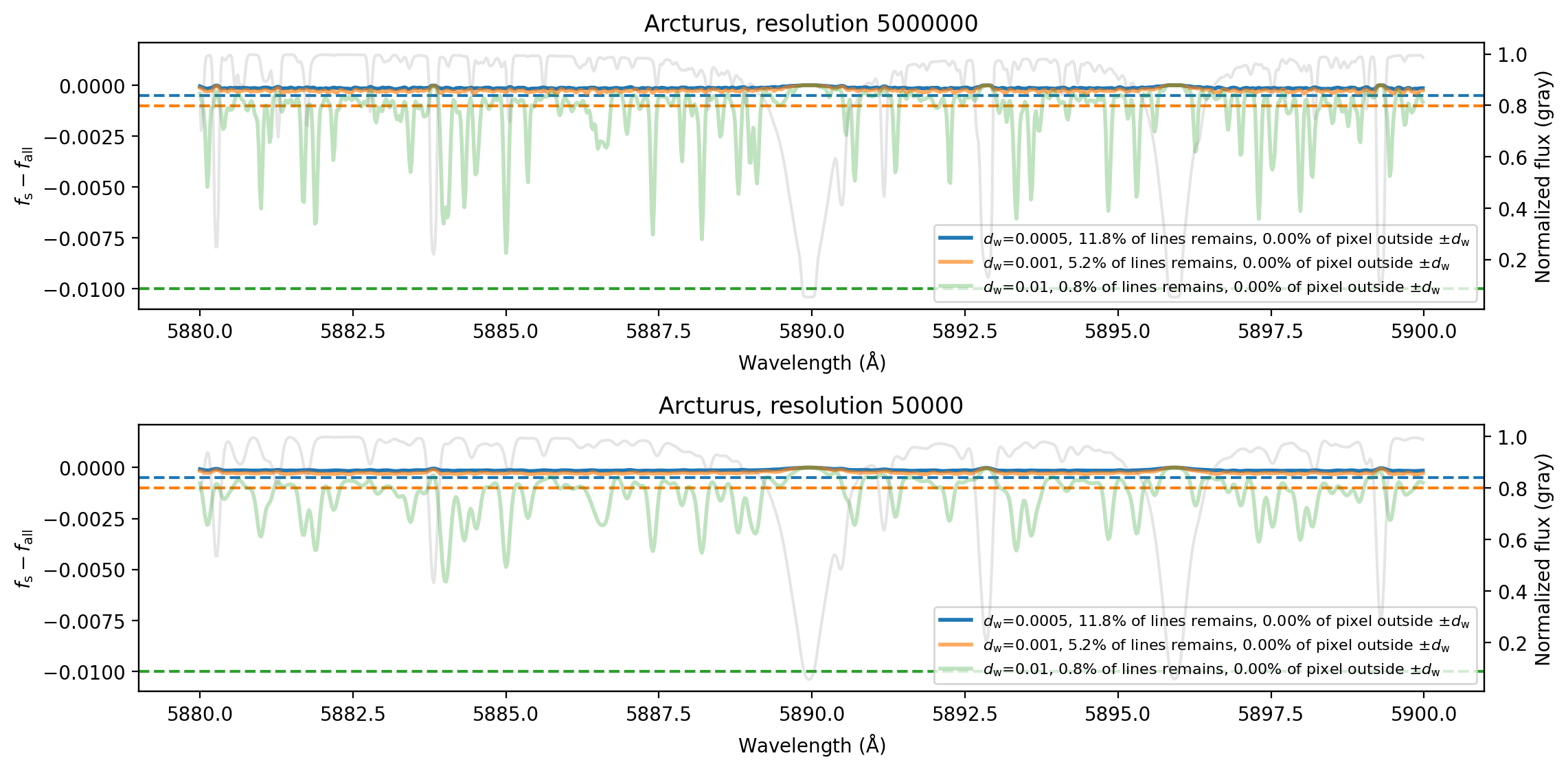}
    \caption{Same as Figure~\ref{fig:cd_thres_sun}, but for the stellar parameters of Arcturus.}
    \label{fig:cd_thres_arcturus}
\end{figure*}

For cool giants, the situation differs due to the enhanced strength of molecular features. 
Figure~\ref{fig:cd_thres_arcturus} presents the corresponding comparison for Arcturus, where a larger fraction of lines (ranging from 0.8\% to 13\%) are kept in the synthesis.
This is primarily driven by the increased strength of TiO and other molecular lines at lower temperatures. Nevertheless, even in this more line-rich regime, the filtered spectra remain within the defined tolerance, with no pixel exceeding the $d_\mathrm{w}$ threshold.
A similar comparison for Arcturus suggests an even larger empirical ALMAX threshold (\(\sim 9\times10^{-3}\)) for \(d_\mathrm{w}=5\times10^{-4}\), reinforcing that the default ALMAX value is conservative in these test cases.

\subsection{$\dc$ across the Kiel diagram}

To extend the validation beyond Sun and Arcturus, we now examine how the fraction of non-negligible lines varies across the stellar parameter space. 
This provides a practical diagnostic for estimating the number of lines that must be retained in a synthesis, depending on the target star's location in the Kiel diagram.

\begin{figure*}
    \centering
    \includegraphics[width=1\linewidth]{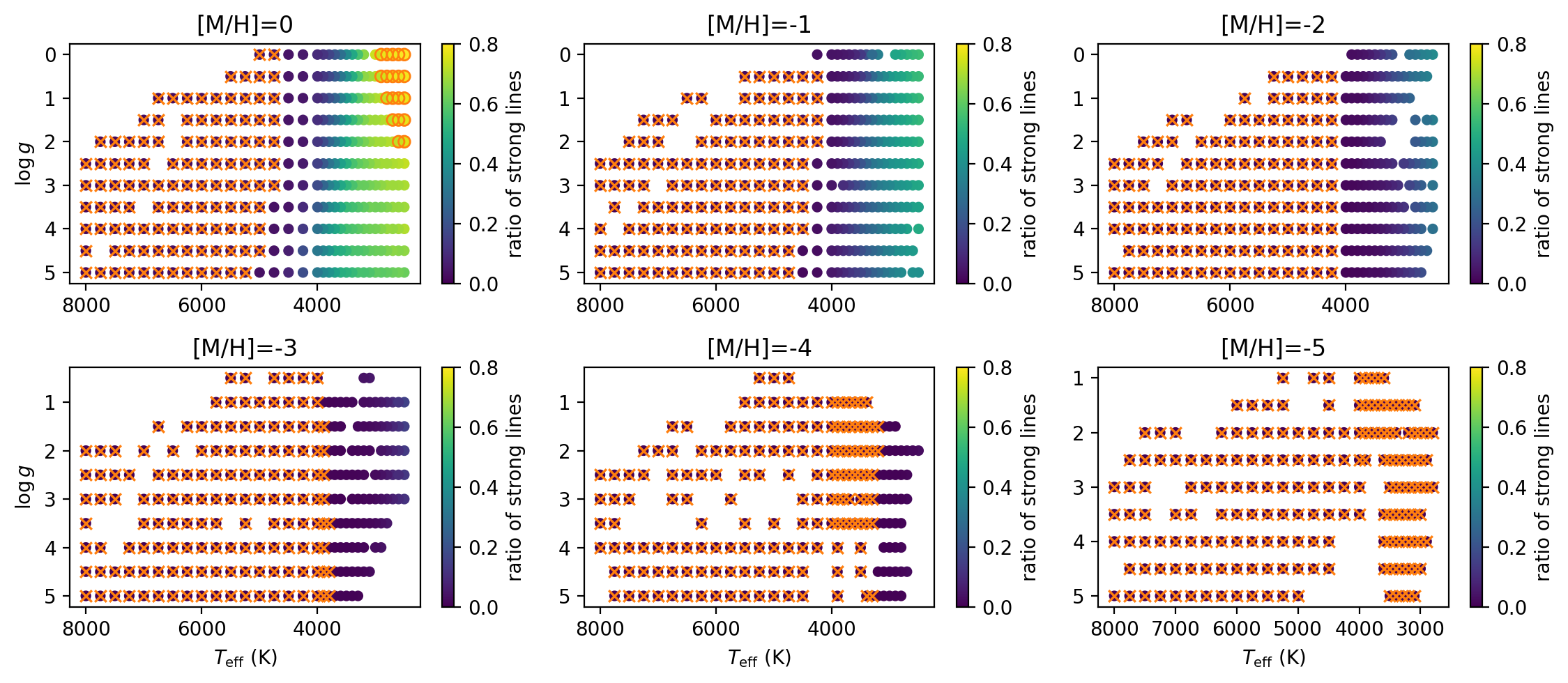}
    \caption{The ratio of non-negligible lines across the Kiel diagram for MARCS grid points. Grid points marked with orange circles indicate locations where more than 75\% of TiO lines are non-negligible, while orange crosses denote points where all TiO lines are negligible.}
    \label{fig:line_cdr_Kiel}
\end{figure*}

We computed $\dc$ and $\lv$ for all lines at each grid point of the MARCS model grid, and classified them as negligible or non-negligible using the method described in Section~\ref{sec:negligible-line-filtering}, adopting a threshold of $d_\mathrm{w}=0.001$.
As shown in Fig.~\ref{fig:line_cdr_Kiel}, for most solar-metallicity stars with $T_\mathrm{eff}>4500\,\mathrm{K}$, fewer than 20\% of lines are non-negligible and thus need to be included in the synthesis.
Toward lower $T_\mathrm{eff}$, the fraction of retained lines increases rapidly, reaching approximately 75\% around $\teff \sim 3000\,$K (and even higher for $\logg\sim0$), where nearly all lines must be included and synthesis becomes significantly slower. 
A similar trend is observed at lower metallicities, although the temperature threshold for retaining more lines shifts to lower values. 
The dependence on surface gravity remains relatively stable across metallicities.
In the extremely metal-poor regime (e.g., $[\mathrm{M}/\mathrm{H}]<-4$), strong lines become increasingly rare or entirely absent, allowing for aggressive line filtering and faster synthesis.

One of the primary factors behind these trends in our sample is the variation in TiO line strength: the rising fraction of strong lines is largely driven by increasingly strong TiO features. 
In some applications it may be more practical to treat TiO molecular lines collectively, either excluding or including all lines, rather than computing the central depth for each transition individually. 
Based on this consideration, we divide the Kiel diagram into three regimes: regions where TiO can be safely ignored, regions where partial inclusion is advisable, and regions where full inclusion is necessary for accurate synthesis.
This classification provides a pragmatic guideline for balancing computational efficiency with spectral completeness in modelling optical spectra of cool stars.

Figure~\ref{fig:line_cdr_Kiel} shows a visual classification of the three regimes introduced above. 
Orange crosses mark grid points with no non-negligible TiO lines, while orange circles indicate regions where more than 75\% of TiO transitions are classified as strong.
At solar metallicity, stars with $T_\mathrm{eff}>5250\,$K have no non-negligible TiO lines, whereas roughly half of the models with $T_\mathrm{eff}=2500$-$3000\,$K and $\log g=0$-$2$ fall in the regime where nearly all TiO lines must be considered. 
At $[\mathrm{M}/\mathrm{H}]=-1$, the negligible region extends to $\sim4500\,$K and the non-negligible region disappears. 
By $[\mathrm{M}/\mathrm{H}]=-5$, the entire Kiel diagram lies in the region where TiO is negligible. 
This map provides a practical guideline for when to include, partially include, or exclude TiO in optical spectral synthesis.

\subsection{Comparison between the old and new \sme}
\label{sec:SMElib_comp}

\begin{figure*}
\resizebox{12cm}{!}{\includegraphics{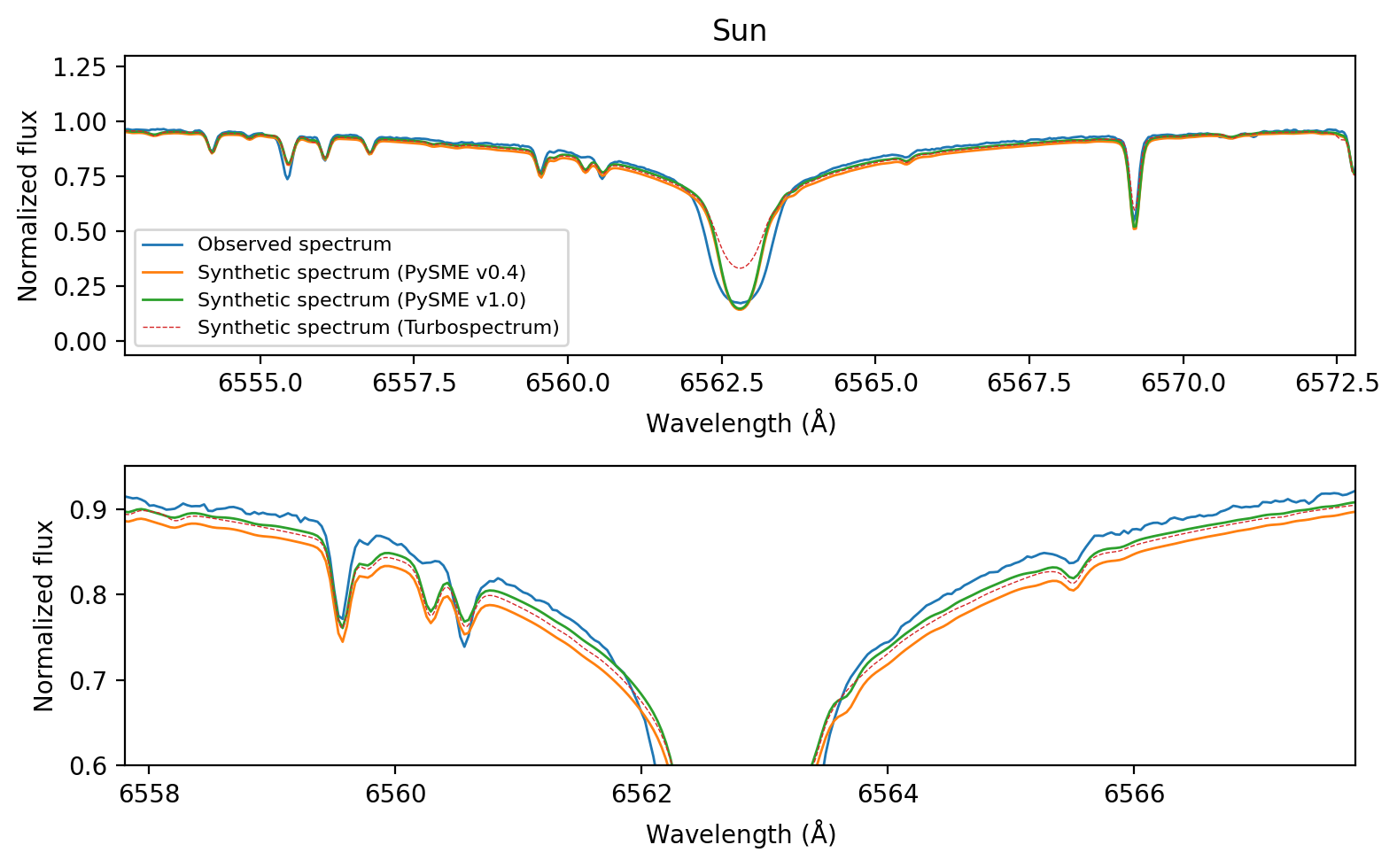}}
\hfill
\parbox[b]{55mm}{
\caption{Comparison between the observed solar spectrum in the H$\alpha$ region and synthetic spectra generated using PySME v0.4, PySME v1.0, and Turbospectrum.}
\label{fig:sun_arcturus_spec}
}
\end{figure*}

The differences between spectra synthesised with \pysme v0.4 (using \sme v6.0) and v1.0 (using \sme v6.13) are generally minor, except for the hydrogen lines. 
Figure~\ref{fig:sun_arcturus_spec} compares the two versions for the Sun in the vicinity of H$\alpha$ and the spectrum generated using Turbospectrum, along with the observed solar spectrum in the Melchiors database \citep{Royer2024}, with a resolution of 85,000. 
The blending features exhibit nearly identical depths and shapes, indicating that the upgrade has minimal impact on most metal lines.
However, the hydrogen equation of state (EOS) has been revised in the new \sme, leading to noticeable changes in the H$\alpha$ wings. 
In particular, the updated version produces higher wing fluxes that align more closely with observed solar spectra, improving the consistency of hydrogen line modelling with observations.
We recommend using the latest version of \sme (and thus the latest \pysme) whenever possible, to ensure optimal agreement with observations and consistency across analyses.

\subsection{Time performance of synthesis under line filtering}

\begin{figure*}
    \centering
    \includegraphics[width=1\linewidth]{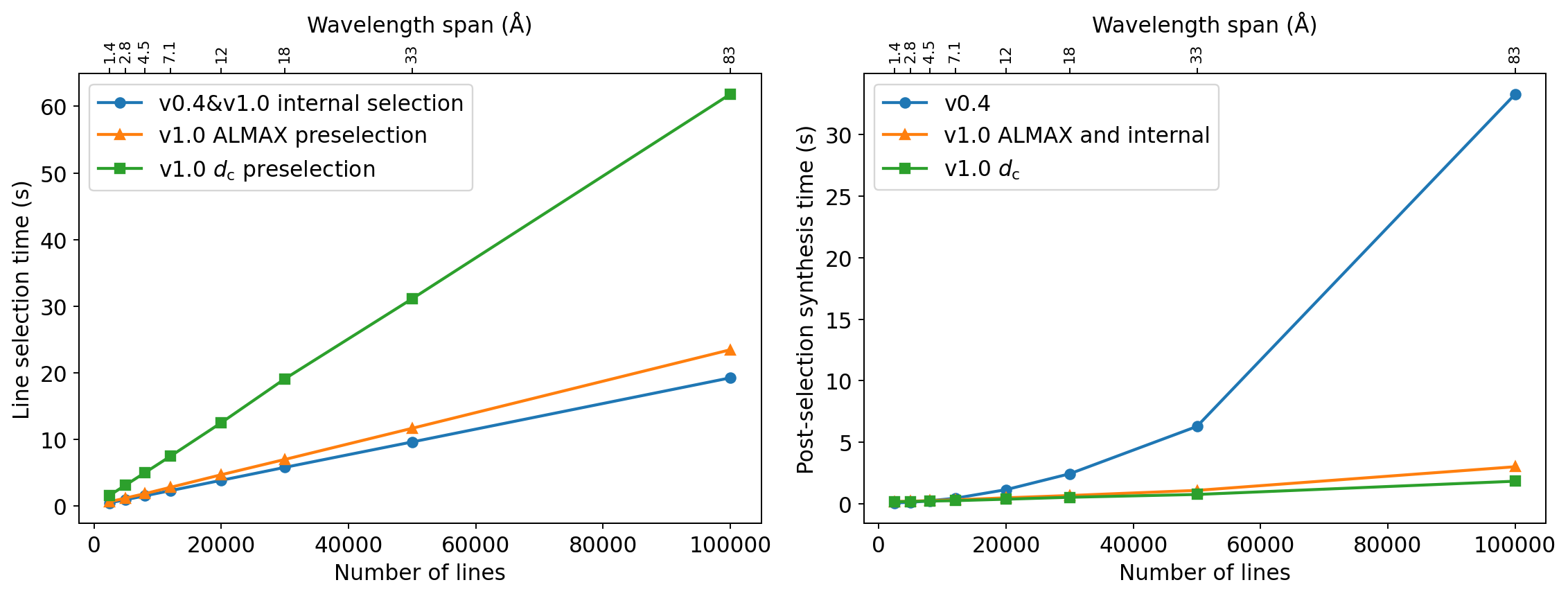}
    \caption{Wall-time comparison of line selection and post-selection synthesis in \pysme\ for synthesis windows of different widths starting at 5880\,\AA.
    Left: serial selection time for the internal (default), ALMAX, and \(d_\mathrm{c}\)-based workflows.
    Right: post-selection synthesis time for \pysme\ v0.4 as well as \pysme\ v1.0 using ALMAX-selected (internal) and \(d_\mathrm{c}\)-selected line lists.}
    \label{fig:synthesize_time}
\end{figure*}

The updated line-selection workflows in \pysme\ shift part of the computational cost from the synthesis stage to an explicit line-selection or preselection step.
Figure~\ref{fig:synthesize_time} separates these two contributions for synthesis windows of different widths starting at 5880\,\AA, using the default thresholds of 0.0001 and 0.001 for ALMAX and $\dc$, respectively.

The left panel shows the time spent on serial line selection.
Among the three workflows, the original selection is the least expensive, ALMAX preselection is moderately more costly, and $d_\mathrm{c}$-based preselection is the most expensive.
Compared with the original workflow, ALMAX preselection introduces additional overhead from the Python side line selection step, while $d_\mathrm{c}$-based preselection is slower still because evaluating the central depth requires a line centre radiative-transfer calculation.
Both ALMAX- and $d_\mathrm{c}$-based preselection can be parallelised, which is a main motivation for introducing these workflows despite their higher serial cost.

The right panel shows the time spent in the subsequent synthesis stage after line selection.
The number of input lines increases approximately linearly from 2,500 to 99,000, while 2\% or 1\% of them are selected as non-negligible lines under ALMAX and $\dc$ selection.
For \pysme\ v0.4, the post-selection synthesis time increases steeply with the number of input lines.
In \pysme\ v1.0, the post-selection synthesis time is substantially reduced over the full range tested here for both the ALMAX- and $d_\mathrm{c}$-based workflows.
Since the original \pysme\ v0.4 design already followed the ALMAX style line selection route, the large gap between \pysme\ v0.4 and \pysme\ v1.0 is primarily due to implementation-level performance improvements in the synthesis stage.
The remaining difference between the ALMAX- and $d_\mathrm{c}$-selected cases mainly reflects the fact that the default $d_\mathrm{c}$ threshold retains fewer lines than the default ALMAX threshold in this test.

In practice, these results highlight a trade-off between preselection cost and interpretability.
Within the Python side workflows, ALMAX is preferable as the default choice when preselection efficiency is the main concern, whereas $d_\mathrm{c}$-based selection is advantageous when a more directly interpretable threshold on line-selection behaviour is desired.
The absolute runtime in Fig.~\ref{fig:synthesize_time} should be interpreted as illustrative rather than universal, as it depends on the adopted ALMAX or $d_\mathrm{w}$ threshold and on the wavelength region being synthesised.

\section{Summary}

In this paper we present \pysme\ v1.0, a Python implementation of Spectroscopy Made Easy built on an updated \sme\ radiative-transfer core.

A key development is a revised framework for line selection, including unified Python-side workflows based on ALMAX and central depth, together with dynamic line list construction and validity-range control.
These workflows build on the existing line-filtering capabilities of \pysme, while making them more flexible, more interpretable, and more scalable for survey-scale applications.
Optional flux contribution functions further visualise the depth of formation of different spectral regions, making the synthesis more interpretable.
The updated equation-of-state treatment improves the modelling of hydrogen lines, particularly Balmer features, while preserving good agreement with previous SME results for metal lines.
The Python wrapper has also been extended to support parameter-dependent derived quantities that can be updated during fitting, so that these diagnostics can be integrated into iterative analysis workflows when needed.
\pysme\ v1.0 further provides NLTE departure-coefficient grids for 17 elements in 1D MARCS atmospheres.

Taken together, the updated physics, flexible line-selection workflows, and improved synthesis efficiency make \pysme\ v1.0 well suited to scalable, precision analyses in large spectroscopic surveys.

\begin{acknowledgements}
M.J. is thankful for the useful discussion with Ross Church and Xiaoting Fu. 
B.T. acknowledges the financial support from the Wenner-Gren Foundation (WGF2022-0041) and from the Royal Physiographic Society in Lund through the Stiftelsen Walter Gyllenbergs and M\"arta och Erik Holmbergs donations.
H.J. acknowledges support from the Swedish Research Council, VR (grant 2024-04989).
This work has made use of the VALD database, operated at Uppsala University, the Institute of Astronomy RAS in Moscow, and the University of Vienna.
\end{acknowledgements}

\bibliographystyle{aa} 
\bibliography{refs} 

\appendix
\section{NLTE Grid References}

Table~\ref{tab:NLTE-grids} lists the references for the model atoms used to compute the NLTE departure coefficient grids implemented in \pysme.

\begin{table}[htbp]
    \centering
    \caption{NLTE model atom references.}
    \label{tab:NLTE-grids}
    \begin{tabular}{ll}
        \hline
        Element & Reference \\
        \hline\hline
        H  & \citet{Zhou2023} \\
        Li & \citet{Lind2013, Wang2021} \\
        C  & \citet{Amarsi2019} \\
        N  & \citet{Amarsi2020} \\
        O  & \citet{Amarsi2018a} \\
        Na & \citet{Lind2011} \\
        Mg & \citet{Osorio2015} \\
        Al & \citet{Nordlander2017} \\
        Si & \citet{Amarsi2017} \\
        S  & \citet{Amarsi2025} \\
        K  & \citet{Reggiani2019} \\
        Ca & \citet{Osorio2019} \\
        Ti & \citet{Mallinson2024} \\
        Mn & \citet{Bergemann2019} \\
        Fe & \citet{Amarsi2022} \\
        Cu & \citet{Caliskan2025} \\
        Ba & \citet{Gallagher2020} \\
        \hline
    \end{tabular}
\end{table}

\end{document}